Corresponding Author: Mr. Alexander Henderson,

Corresponding Author's Institution: Rice University

First Author: Alexander Henderson

Order of Authors: Alexander Henderson; Edison Liang; Nathan Riley; Pablo Yepes; Gillis Dyer; Kristina Serratto; Petr Shagin

Ultra-Intense Gamma-Rays Created Using the Texas Petawatt Laser

Abstract:

In a series of experiments at the Texas Petawatt Laser (TPW) in Austin, TX, we have used attenuation spectrometers, dosimeters, and a new Forward Compton Electron Spectrometer (FCES) to measure and characterize the angular distribution, fluence, and energy spectrum of the X-rays and gamma rays produced by the TPW striking multi-millimeter thick gold targets. Our results represent the first such measurements at laser intensities $> 10^{21}$ W*cm$^{-2}$ and pulse durations < 150 fs. We obtain a maximum yield of X-ray and gamma ray energy with respect to laser energy of 4% and a mean yield of 2%. We futher obtain a Full Width Half Maximum (FWHM) of the gamma distribution of 37°. We were able to characterize the gamma-ray spectrum from 3 MeV to 90 MeV using a Forward Compton Electron Spectrometer, with an energy resolution of 0.5 MeV and mean kT of ~ 6 MeV.. We were able to characterize the spectrum from 1 to 7 MeV using a Filter Stack Spectrometer, measuring a mean gamma-ray temperature for the spectrum from 3 to 7 MeV of 2.1 MeV.

I: Introduction

When a laser of intensity greater than a few times $10^{18}$ W/cm$^2$ strikes a solid, high-Z target it couples a significant fraction of its energy to hot electrons with kinetic energy > $m_e c^2$.[1,2, 3,4] In a dense, high-Z, conductive material such as gold, a significant fraction of these electrons will undergo bremsstrahlung, producing copious gamma and x-rays. See Gibbon 2005 [5] for a review of this topic. The development in recent years of petawatt-class lasers such as the Texas Petawatt Laser (TPW) in Austin, TX opens up new potential for producing and studying ultra-intense, short-pulse multi-MeV gamma rays.

In petawatt class lasers, as much as 50% of the laser energy can be converted to hot electrons [6]. In principle, based on our Monte Carlo simulations, 30-50% of hot electron energy is convertible to gamma-rays, for a total laser to gamma-ray conversion factor of 15-25%, far in excess of any other mechanism. We measure a maximum efficiency of laser energy to gamma rays of 4% and a mean conversion efficiency of 2%. This would result in a lower laser energy to hot electron energy conversion rate of about 5-15%. The total laser-energy-to-gamma-ray conversion efficiency we obtain is short of the theoretical maximum but still high enough to be promising.

Intense gamma-ray beams have potential novel applications for experimental physics and new technologies. For instance, colliding intense gamma-ray beams can potentially create a pure pair plasma via (γ,γ) pair production if the gamma-ray fluence is above a certain threshhold. [7] For other pair-production mechanisms such as Trident and Bethe-Heitler [7], the e+/e- ratio is far short of a true pair plasma [8]. Such gamma-ray beams may also be useful for interrogating material hidden inside shielded containers, i.e. detecting concealed nuclear materials, and for nuclear medicine.

In the process of attempting to measure and characterize the gamma-ray beams produced by the interaction of petawatt class lasers with high-Z solid targets, we have also applied a method of gamma-ray spectrum measurement using Compton scatting, as suggested by Ahrens et al.[9] in 1972, developed by Matscheko and Ribberfors[10] and Morgan et al. [11], and more recently used by such groups as Reims et al. [12]. Following the example of Morgan et al. [11], we have chosen to consider Compton electrons scattered near 0 degrees. Then, we measure a wide range of Compton electron energies and adjust the acceptance angle and the thickness and type of converter material in order to sample different energy ranges. We refer to this method hereafter as Forward Compton Electron Spectroscopy (FCES). We measured the energy of the electrons using a rare earth permanent magnet with peak field strength of about 0.6 Gauss which had previously been modeled using GEANT4 Monte Carlo simulation [13, 14] based on a measured field map and calibrated by using mono-energetic electron beams from a linear accelerator. We refer to this device hereafter as a Forward Compton Electron Spectrometer (FCES). In our experiments, we achieved good results with a combination plastic and copper converter and with Compton electrons accepted into a ~ 5˚ angle, with the deconvolved spectrum matching Monte Carlo models between 3 and 50 MeV.

As in the method proposed by Morgan et al. [11] and Reims et al. [12], this method of measuring gamma ray spectra using Compton scattering and permanent magnets is compact, light, and simple enough that it could potentially be deployed to measure the energy spectrum of high-energy medical x-ray lines with greater ease and efficiency than the methods commonly used today [12].

II: Experimental Set-up

Experiments were carried out at the TPW located in Austin, TX in Target Chamber 1 (TC1) for high-intensity shots with an f/3 dielectric parabolic mirror donated by Los Alamos National Lab using gold targets. The cylindrical targets had thicknesses from 0.2 mm to 1 cm and diameters from 2 mm to 4.5 mm. The angle between target normal and laser forward was varied between 25 degrees and 45 degrees. The laser and target parameters are listed in Table 1.

Magnetic spectrometers were used to measure the electron and positron spectra as well as proton spectra from back-surface contaminants. In addition, two Filter Stack (or attenuation) Spectrometers (FSS) (see Section IV) were employed to measure the low-energy gamma ray spectrum, and a FCES was used on some runs to measure the gamma spectrum at higher energies (Section V). The full layout of TC1 is shown in figure 1.

We used FLUKA, a particle transport code, to model the FSS's response and GRAVEL, a spectrum unfolding code which is built based on the SAND-II code, is used to take the FSS image plate scans and obtain X-ray spectra. GEANT4, a Monte-Carlo code, was used to model the response of the FCES as well as to model the creation of high-energy gamma rays by hot electrons incident on gold targets. [13,14]

The measured emergent electron spectrum for a 0.35 mm target (Fig. 2) was used as the hot electron spectrum for the Geant4 Monte-Carlo simulation. The measured spectrum has excess attenuation at lower energies compared to the true incident spectrum, but for energies above 1.5 MeV this should not be significant. The angular distribution of incident electrons was assumed to be a Gaussian with a Full Width, Half Maximum (FWHM) of 20 degrees. This was based on the work of Chen et al. [15] and is consistent with the results of Dosimeter measurements of gamma and x-rays produced by the electrons (Section VI). The time profile was assumed to be a Gaussian with a FWHM of 150 fs, the duration of the main pulse of the TPW.

Section III: Filter Stack Spectrometer Results:

Two filter stack X-ray spectrometers were used to characterize the Bremsstrahlung spectrum, based on work by C. D. Chen et al.[16] Due to the high X-ray fluence produced at the target, the spectrometer design chosen relies on differential filtering with high-Z absorbers. The detector media are image plates, which are removed and read out after every shot. Figure 3 shows a diagram of one of these spectrometers. Either seven or eleven channels (depending on spectrometer) are read out directly, and numerical deconvolution is used to produce the recorded energy spectrum. Measured data is recorded as photostimulated luminescence (PSL), proportional to deposited photon energy. This proportionality is effectively linear over the energy range of interest.

In order to deconvolve the spectra, the detector response is modeled numerically using the FLUKA particle transport code.[17, 18] A sequence of ten monoenergetic photon beams per energy decade was propagated through the filter stack, and the resulting energy deposited in the image plate layers was recorded. The output spectrum is calculated using the GRAVEL code, which uses a modification of the SAND-II regularized nonlinear least squares algorithm.[19, 20] The resulting output spectrum was binned in regions corresponding to the monoenergetic response function inputs.

C. D. Chen et al. [21] show, in their figure 2, that for photons of up to about 5 MeV, the energy deposited in at least one of the IP layers by that photon will differ from the energy deposited in the other IP layers by that same photon. Thus, it should in principle be possible to use this sort of spectrometer to measure photon energies up to 5 MeV through careful analysis.

To calibrate this instrument we used an X-ray beam line with a known spectrum at the Mary Bird Perkins Cancer Center (MBPCC) in Baton Rouge, Louisiana. In figure 4, we can see that the deconvolved FSS spectrum agrees with the input X-ray spectrum within 30% between 1 and 2.5 MeV, and within 10% between 3 and 5 MeV. Below around 1 MeV the spectrum is highly inaccurate as a result of the materials used: use of thinner layers of the low-Z materials would provide better results in this range. We may thus compare the FSS results on TPW experiments with simulation results in the range of 1 to 5 MeV. This compares favorably with the results of C. D. Chen et al. [21]

Figure 5 shows results from the TPW experiments using the FSS and comparisons with our Geant4 Monte Carlo simulation. We compare the Monte Carlo spectrum obtained from our hot electron spectrum passing through different Au target thicknesses with the results of the FSS for that target and achieve a good match, considering the accuracies mentioned above, with the exception of an occasional "bump" between 1 and 3 MeV that is likely due to artificial numerical effects in the unfolding algorithm, since it shows up in some runs measuring a known, "smooth" spectrum.

From the exponential "tail" of the FSS spectrum in a log-linear plot (Figure 6 a-b) we obtain effective temperatures (kT) of about 1.6 to 2.9 MeV for the Gamma ray distribution, with a mean of 2.1 MeV. The hot input electrons had a temperature of about 16 MeV, so we suspect that in this energy range (3 -7 MeV), bremsstrahlung gamma rays from secondary electrons dominate over those from primary electrons. Figure 6c shows an overlay of GEANT4 Monte-Carlo simulation results showing a temperature of around 2.5 MeV, within the measured range. In Figure 6d the low energy gamma temperature is roughly inversely correlated with laser energy, but Figure 6e does not show any apparent correlation between laser intensity and gamma temperature.

Section IV: Forward Compton Electron Spectrometer Results

The gamma ray spectrum with energies greater than a few MeV cannot be measured effectively using the FSS due to the energy-independence of the attenuation coefficient. Yet the short pulse nature of the TPW-produced gamma rays also cannot be measured using single-photon counting detectors such as scintillators. Hence we decided to employ a technique suggested by Morgan et al. [10]. When gamma-rays hit a low-Z thin target, the Compton scattered electrons in the forward direction (near 0°) have an energy which correlates simply with the the generating gamma ray through the Compton scattering formula in the small angle approximation about 180º [22]:

$$\lambda' - \lambda = \frac{h}{m_e c}(2 - \frac{\theta^2}{2})$$

where $\theta$ is the angle of the electron from photon incidence, or alternately the deviation of the scattered photon from 180º from photon incidence.

The idea, then, is to use these forward Compton scattered electrons to measure the high-energy gamma ray spectrum by narrowly collimating both the incident gamma-rays and scattered electrons.

For our FCES (Fig. 7), we use a magnetic electron spectrometer which consists of image plates on both sides of a gap between rare earth magnets. We then attached a Compton scatterer, consisting of varying combinations of polyethylene, aluminum, copper, and tin plates. In front of this we inserted a narrow collimator made of copper and lead totaling over 10 cm with a 6 mm diameter hole to collimate gamma rays. The FCES is placed nearly flush with an external glass porthole of TC1 (Fig. 1) at a distance of approximately 1 m from the target, so that the incident gamma ray beam is < 1° from the converter normal direction. The acceptance angle of Compton electrons was varied by means of teflon blocks with 6 mm diameter holes in them, so that the acceptance angle could be varied from 25° to 4.5°. The image plates were placed on both sides of the magnet gap to catch both positrons, electrons, and scattered x-rays, thus providing a way to subtract out external background and pair creation electrons, which compete with Compton electrons for thick converters. However, the acceptance angle was wide enough that electrons were distributed across the magnet gap, so an unknown number of electrons impacted the magnet. Thus we cannot gauge the absolute magnitude. However, we can still track the relative magnitude and obtain a spectral shape.

In order to properly deconvolve the FCES plate images, we simulate the predicted Compton electron spectrum using GEANT4. In this simulation, we shoot a large number of gamma rays at fixed energy increments, e.g, 10000 at 0.5 MeV, 10000 at 1.0 MeV, etc. The resulting point-spread function (psf) can be inverted to obtain the original gamma ray distribution given the distribution of Compton and pair-produced electrons. While this specific simulation was not directly validated, similar simulations have been validated for the production of gamma rays (see section III above), as well as positrons and electrons under different conditions [23, 24].

Figure 8 shows the results of an end-to-end simulation, from photon incidence to the electron registering on the image plate. In practice this was not how the images were deconvolved, but the results are illustrative. We also observe that while there are electrons all along the plate above a certain point, there are none above it. This corresponds to the fact that Compton scattering produces a continuum of electrons with energies less than the energy of the photon. We can then use the spacing of this cut-off between two photon energies to determine an approximate energy resolution. The image plates are resolvable by our instruments to a size of 100 microns. This spacing in electron cut-offs corresponds to a difference in photon energy of about 0.1 MeV. It varies along the length of the magnet, so we use 0.5 MeV energy bins in simulation and take that as our energy resolution in order not to under-estimate it.

Next, we constructed a detector response matrix (DRM) of the gamma-ray-to-Compton-electron process and used a previously developed psf [23] for the electron-to-image-plate signal to deconvolve the image plate results into a gamma-ray spectrum. These psfs were constructed by simulating a large number ($10^3$) of particles injected at fixed energies in 0.5 MeV intervals and using Matlab to obtain a pseudo-inverse of the resulting matrix. In order to remove as much noise and non-Compton electron background as possible, first the images are separated into a "background" region, and a "signal" region. Then, the signal level on each image is integrated along the y-axis and the background level determined by subtracting the positron image. This removes the effects of pair production and, for the most part, brehmstrahlung scattering in the spectrometer since these processes will affect both positron and electron sides roughly evenly, as seen in Figure 8. Importantly, while many more gamma rays hit the plate than electrons, each electron produces approximately one hundred times the activation in the plate [25]. Then, the averaged value of the background regions is subtracted from each corresponding point in the signal region along the direction of changing energy. See Figure 9 for a flow-chart of the process.

Figure 10 show sample results of deconvolving the spectra for several shots of 1 mm thick gold targets. The electron spectrometer was limited to electron energies <= 50 MeV, so gamma rays >= 50 MeV could not be measured accurately. The region below 1.3 MeV is similarly affected. In addition, electron energy such that the mean free path (mfp) is less than the thickness of the converter (7 mm) is 10 MeV for tin and 7 MeV for copper, so emergent scattered Compton electrons whose energy is less than these energies provide only a partial sampling of gamma rays below those energies. We compare the remaining region with simulation. We find that for a converter consisting of 12.5 mm of plastic and 7 mm of copper we are able to accurately characterize the gamma-ray spectrum from 3 MeV to 50 MeV, with an energy resolution of approximately 0.1 MeV and up, taken as 0.5 MeV, as noted above. Ahren et al [9] proposed a detector on a similar principle to measure up to 300 MeV with a resolution of 20 keV at 22 MeV. Matschekot and Ribberfors produced a Compton spectrometer using thin lucite, a material with a much lower Z-value and density than we used, that functioned at 20-200 keV with a resolution of 0.42 keV. [10] With a more clever magnet arrangement or a longer track of weaker magnets we may be able to achieve similar resolutions, though at an increased difficulty in fabrication.

From the exponential tail of the spectrum, an effective temperature (kT) may be determined. From our TPW data, we obtain a mean temperature of 6.6 MeV. Note that this is significantly higher than the 2.1 MeV mean temperature seen using the FSS spectrometer, supporting the conclusion in section IV that the FSS spectrum was dominated by bremsstrahlung from secondary electrons. Figure 10d shows the GEANT4 Monte-Carlo simulation for the gamma rays in the appropriate region. This simulation shows the gamma rays having a temperature of about 7.2 MeV, a little higher than the data from the FCES. The difference is likely due to there being some variation in the incident electron temperature, while the simulation consistently used the electron spectrum from the thinnest gold shot. In all cases the temperature of the gamma rays is lower than that of the electrons. This is most likely due to the scattering processes generating the gamma rays.

Section V: Dosimeter Angular Distribution and Total Gamma Yield

We measured the angular distributions of the daily-cumulative gamma rays by using Landauer Pa area dosimeters placed on the external surface of TC1's tank's wall to measure the dose at ~1 m away from the target. The energy response range of the dosimeters is 5 keV to 40 MeV and a readable dosage range from 1 mrem to 1000 rem. This experiment was carried out during our runs at the Texas Petawatt Laser in 2011, 2012, and 2013. 23 total data sets were collected, but owing to internal instruments obstructing many dosimeter directions on some shots, only 20 sets were usable. These included 4 data sets where platinum targets were used instead of gold, 4 where a mix of platinum and gold targets were used, and 5 sets where differing target shapes such as long thin rods, targets with conic tips, and targets with concave or slant surfaces were used. We exclude the platinum in our discussion since there are not enough of them to examine collectively, and we also exclude the mixed sets since the effects of this mix are not easily determined. Different data sets had different angles between laser forward direction and target normal direction, and each data set is the accumulated dosage of between 2 and 8 laser shots.

The angular distributions were fitted to Gaussian distributions and then integrated over the sphere to obtain a total dose, which was then converted into a total gamma energy. We also account for the attenuation from the chamber walls (~1 cm of steel) to obtain a total emergent gamma energy. We accounted for the attenuation by using the simulated spectrum and NIST attenuation data [26]. The results are compiled in Table 2. The conversion efficiency from incident laser energy to gamma rays for the 100 J laser configuration varies from 0.94% to 3.8%, with a mean efficiency of around 2 %. The two days with efficiencies in excess of 3% used targets that were not flat disks or rods. Specifically, they used a combination of "lense" targets, where the back of the target (opposite the laser incidence) and "slant rod" targets, where the surface of the rod facing laser incidence was slanted such that the length of the rod could run down the direction of the laser without reflected laser light damaging the laser apparatus.

Figure 11 shows the gamma-ray yield for gold targets versus laser fluence and versus laser energy. While there is no clear correlation between laser fluence and yield, there is a roughly quadratic relationship between laser energy and gamma ray yield up to around 100 J. Above 100 J there is no clear correlation.. There are two significant outliers in Figure 11b: the one at about 45 J corresponds to the data taken in 2011, before the laser was upgraded, and the point at 114 J and 0.9% efficieny corresponds to a dosimeter set when the laser energy was pushed beyond its recommended safety tolerances on the first shot of the day, which may have affected beam quality throughout the day.

Hatchett et al. [27] measured a mean total x-ray energy of 11 J from the Petawatt laser at Lawrence Livermore National Laboratory, whose energy varied from 150-750 J, averaging around 375 J, with an intensity of $3 \times 10^{20}$ W/cm$^2$ on average, incident on 50 – 125 micron thick gold targets. This gives a mean conversion efficiency of about 3%, comparing favorably with our results. The conversion efficiency of the 50 J laser (the shots in 2011) to gamma rays was only 0.78%.

Figure 12 shows several angular distribution results, while figure 13 shows similar results in polar form. The Full Width, Half Maximum (FWHM) of the Gaussian distribution was about 17° for laser energies of around 45 J, while for more typical energies of ~100 J - 120 J the FWHM ranged from 13° to 53°, with a mean of 33°. This compares favorably with the results obtained by Norreys et al of 35°. [28] GEANT4 simulation shows a FWHM of 13° (figure 14), but this is using an incident electron distribution with a FWHM of 20°. This was based on the measured electron distribution of the 50 J configuration of TPW, and indeed the simulated gamma-ray angular distribution is very close to that measured for that laser configuration. The generally broader distribution for 2012 and 2013 indicates that the electron distribution for the 100 J laser was often much broader, with a FWHM sometimes exceeding 40°. This seems to depend partly on the angle between laser forward and target normal and partly on the shape of the target. While it is not obvious in the figures, Table 2 shows that the mean angle between the peak of the gamma-ray distribution and the target normal was 24°, towards laser forward, which is close to the PIC simulation results of Sheng et al. [29] of 30° for gamma rays of energies between 0.5 and 30 MeV.

Section VI: Conclusions

For a petawatt class laser irradiating millimeter-thick gold targets, we were able to characterize the gamma-ray spectrum from 3 MeV to 90 MeV using a Forward Compton Electron Spectrometer, with an energy resolution of 0.5 MeV and mean kT of ~ 6 MeV.. We were able to characterize the spectrum from 1 to 7 MeV using a Filter Stack Spectrometer, measuring a mean gamma-ray temperature for the spectrum from 3 to 7 MeV as 2.1 MeV. Dosimeter results show a mean conversion efficiency from laser energy to gamma rays of around 2%, a mean angle between the peak of the gamma distribution and target normal of 26°, and a FWHM of the gamma distribution of 37°.

GEANT4 Monte-Carlo simulation has proven to be adequate for describing the gamma-ray angular distribution and spectrum as well as deconvolving and testing the results from a Forward Compton Electron Spectrometer. In the future we intend to couple this simulation with a PIC code to get an accurate representation of the incident hot electron spectrum and angular distribution and thus to describe the center of the gamma-ray angular distribution with respect to target normal and laser forward.

In addition, we hope to extend the range of the spectrum measured for both the FSS and FCES using more detailed simulation and careful background subtraction, as well as different strengths and lengths of magnets for the FCES. We also plan to make further use of dosimeters and simulation to characterize and predict the gamma-ray angular distribution and flux. Once these techniques have been refined, we hope to apply them to measuring the spectra of medical x-ray cancer treatment devices.

We also plan to test, through both simulation and experimentation, ways to optimize the gamma-ray yield and angular distribution and ways of using these gamma rays to interrogate shielded nuclear materials.

Figure Captions:

Figure 1a: Overhead Diagram of Target Chamber 1 (TC1) at the Texas Petawatt laser.

Figure 1b: Simple Schematic of TC1 (not to scale). The radius of TC1 was 1 meter. The diagram shows sample positions of the Filter Stack Spectrometer (FSS) and magnetic spectrometers, and the position of the Forward Compton Electron Spectrometer (FCES). When the FCES was in use, the FSS and magnetic spectrometer along its line of sight were moved to a slightly wider angle.

Figure 2a: Electron energy spectrum used for GEANT4 simulation input. The spectrum is the measured output spectrum for a 1 mm gold target shot on the Texas Petawatt Laser.

Figure 2b: Electron angular distribution used for GEANT4 simulation. The Full Width Half Maximum is approximated from dosimeter data from Texas Petawatt Laser experiments.

Figure 3: Filter Stack Spectrometer (FSS) diagram showing type, placement, and relative thickness of materials for one of the two used. The other spectrometer used had 11 channels instead of 7, using the same materials but with additional thicknesses.

Figure 4: Filter Stack Spectrometer results from calibrating against a known x-ray source. The approximately known source spectrum is represented by circles and the measured spectrum is represented by triangles.

Figure 5a: Filter Stack Spectrometer (FSS) results from the Texas Petawatt Laser. The red (solid) line is FSS data and the blue (jagged) line is GEANT4 Monte-Carlo simulation. The target thickness was 1 mm, the laser energy on the target was 100 J and the intensity was $4.3*10^{20}$ W/cm$^2$.

Figure 5b:  Filter Stack Spectrometer (FSS) results from the TexasPetawatt Laser.  The red (solid) line is FSS data and the blue (jagged) line is GEANT4 Monte-Carlo simulation. The target thickness was 2 mm, the laser energy on the target was 114 J and the intensity was $7.5*10^{20}$ W/cm$^2$.

Figure 5c:  Filter Stack Spectrometer (FSS) results from the Texas Petawatt Laser.  The red (solid) line is FSS data and the blue (jagged) line is GEANT4 Monte-Carlo simulation. The target thickness was 1 mm, the laser energy on the target was 103 J and the intensity was $3.57*10^{20}$ W/cm$^2$.

Figure 6a:  Filter Stack Spectrometer tail from the Texas Petawatt Laser.  The target thickness was 0.5 mm, the laser energy on the target was 102 J and the intensity was $3.96*10^{20}$ W/cm$^2$.

Figure 6b:  Filter Stack Spectrometer tail from the Texas Petawatt Laser.  The target thickness was 1 mm, the laser energy on the target was 100 J and the intensity was $4.3*10^{20}$ W/cm$^2$.

Figure 6c:  Filter Stack Spectrometer tail from the Texas Petawatt Laser seen in Figure 6b, with the corresponding results from a GEANT4 Monte-Carlo simulation overlayed (jagged line).

Figure 6d:  Laser energy versus temperature for the Filter Stack Spectrometer, showing a rough inverse correlation.

Figure 6e:  Laser intensity versus temperature for the Filter Stack Spectrometer, showing no obvious correlation.

Figure 7:  Forward Compton Electron Spectrometer (FCES) diagram, side view.

Figure 8a:  Results of a GEANT4 simulation of the Forward Compton Electron Spectrometer (FCES) response to monoenergetic 15 MeV gamma rays showing the gamma and x-ray background seen after subtracting the gamma ray spectrum on the positron side of the spectrometer from the gamma ray spectrum on the electron side and adjusting for the face that gamma rays activate image plates only about 1/100th as much as electrons.

Figure 8b:  Results of a GEANT4 simulation of the Forward Compton Electron Spectrometer (FCES) response to monoenergetic 15 MeV gamma rays showing an electron signal versus distance along the magnet.  This signal is a result of the electron signal on one side of the magnet minus the positron signal on the other side, thus removing the effects of pair production.

Figure 8c:  Results of a GEANT4 simulation of the Forward Compton Electron Spectrometer (FCES) response to monoenergetic 35 MeV gamma rays showing an electron signal versus distance along the magnet.  This signal is a result of the electron signal on one side of the magnet minus the positron signal on the other side, thus removing the effects of pair production.  Note that the spectral shape and peak position are different than they are with 15 MeV gammas incident.

Figure 9:  Flow-chart demonstrating the process for turning an image plate read-out into a gamma-ray spectrum.

Figure 10a:  Comparison between Forward Compton Electron Spectrometer (FCES) deconvolved spectrum (red) and simulation (blue).  The FCES used a converter consisting of 12.5 mm plastic followed by 7 mm of copper.  A teflon collimator was used to reduce the acceptance angle to 5.5 degrees.

Figure 10b:  Comparison between Forward Compton Electron Spectrometer (FCES) deconvolved spectrum (red) and simulation (blue).  The FCES used a converter consisting of 12.5 mm plastic followed by 7 mm of tin.  A teflon collimator was used to reduce the acceptance angle to 5.5 degrees.

Figure 10c:  Comparison between Forward Compton Electron Spectrometer (FCES) deconvolved spectrum (red) and simulation (blue).  The FCES used a converter consisting of 12.5 mm plastic followed by 7 mm of copper.  A teflon collimator was used to reduce the acceptance angle to 11.5 degrees.

Figure 10d:  Measurement of the slope of the gamma ray spectrum in GEANT4 simulation.  Only the energy range 6-50 MeV is considered, in keeping with the limitations of the FCES.  The target used was a 1 mm thick gold target, and the electron spectrum was as seen in figure 2.

Figure 11a: Dosimeter results for gold targets showing laser fluence versus efficiency.  While the 2011 data, with an average fluence below $10^{20}$ W/cm$^2$, does show a lower efficiency than the data sets with fluences at least a few times $10^{20}$ W/cm$^2$, the efficiency is nearly flat in fluence otherwise.

Figure 11b: Dosimeter results for gold targets showing laser energy versus efficiency. The efficiency increases roughly exponentially, except for one point below 1% efficiency but above 110 J of laser energy, where the laser was pushed past its recommended safety tolerances, which may have affected the beam quality.

Figure 12a: Dosimeter results for 2012, day 5, which had 4 lasers shots with energy of about 100 J each and which had an angle between laser forward and target normal of 35 degrees. All targets were gold.

Figure 12b: Dosimeter results for 2012, day 6, which had 5 lasers shots with energy of about 100 J each and which had an angle between laser forward and target normal of 45 degrees. All targets were gold.

Figure 12c: Dosimeter results for 2013, day 7, which had 7 lasers shots with energy of about 108 J each and which had an angle between laser forward and target normal of 25 degrees . All targets were gold targets with thicknesses much greater than the target diameter ("rod targets").

Figure 13a: Dosimeter polar plot for 2012, day 5, which had 4 lasers shots with energy of about 100 J each and which had an angle between laser forward and target normal of 35 degrees. All targets were gold.

Figure 13b: Dosimeter polar plot for 2012, day 6, which had 5 lasers shots with energy of about 100 J each and which had an angle between laser forward and target normal of 45 degrees. All targets were gold.

Figure 13c: Dosimeter polar plot for 2013, day 7, which had 7 lasers shots with energy of about 108 J each and which had an angle between laser forward and target normal of 25 degrees. All targets were gold targets with thicknesses much greater than the target diameter ("rod targets").

Figure 14: Gamma-ray angular distribution from GEANT4 simulation, using a 20 degree Full-Width-Half-Maximum electron distribution. The relative lack of gamma rays near 0 degrees results from gamma rays being produced by the scattering of relativistic electrons, resulting in the gamma rays having small but non-zero deflection.

**Table 1**

| Year | 2011 | 2012 | 2013 |
|---|---|---|---|
| Total Laser Energy (J) | 37.1 - 48.4 | 82 - 130 | 86 - 129 |
| Pulse Duration (fs) | 182-433 | 141 - 245 | 120-352 |
| Peak Laser Power (TW) | Not Available | 450 - 802 | 326-899 |
| Peak Laser Intensity (W/cm$^2$) | $1.6*10^{19}$ - $2*10^{20}$ | $2.83*10^{20}$ – $1.85*10^{21}$ | $1.9*10^{20}$ – $1.1*10^{21}$ |
| Mean Laser Energy (J) | 43.8 | 98 | 109 |
| Mean Pulse Duration (fs) | 330 | 169 | 160 |
| Mean Peak Laser Power (TW) | Not Available | 585 | 689 |
| Mean Peak Laser Intensity (W/cm$^2$) | $1.0*10^{20}$ | $8.3*10^{20}$ | $5.4*10^{20}$ |

Table 1: Texas Petawatt Laser (TPW) parameters. A major upgrade was performed on the TPW between 2011 and 2012.

**Table 2**

| Year, Day | Total Laser Energy (J) | Total Gamma Energy (J) | Number of Shots | Average Laser Energy Per Shot (J) | Angle between target normal and laser forward (Degrees) | Angle Between Target Normal And Peak of Gamma Distribution (Degrees) | FWHM of Gamma Distribution (Degrees) | Efficiency (%) | Target Type |
|---|---|---|---|---|---|---|---|---|---|
| 2011 | 219 | 1.70 | 5 | 43.5 | 25 | 10 | 17 | 0.78 | Au |
| 2012, 1 | 477 | 6.74 | 5 | 95.4 | 25 | 35 | 19 | 1.4 | Au |
| 2012, 2 | 341 | 3.22 | 3 | 114 | 25 | 21 | 13 | 0.94 | Au |
| 2012, 5 | 404 | 7.69 | 4 | 101 | 35 | 31 | 31 | 1.9 | Au |
| 2012, 6 | 510 | 10.8 | 5 | 102 | 45 | 25 | 28 | 2.1 | Au |
| 2012, 7 | 602 | 9.04 | 6 | 100 | 40 | 21 | 26 | 1.5 | Au |
| 2012, 8 | 786 | 12.6 | 8 | 98.3 | 40 | 24 | 36 | 1.6 | Au |
| 2012, 10 | 730.7 | 9.97 | 8 | 91.3 | 35 | 29 | 39 | 1.4 | Au |
| 2013, 5 | 521 | 20.0 | 5 | 104 | 25 | 18 | 43 | 3.8 | Au |
| 2013, 6 | 574 | 21.3 | 5 | 115 | 25 | 12 | 53 | 3.7 | Au |
| 2013, 7 | 758 | 11.1 | 7 | 108 | 25 | 28 | 46 | 1.5 | Au |
| 2013, 9 | 866 | 24.1 | 7 | 123 | 25 | 24 | 29 | 2.8 | Au |

Table 2: Summary of Au dosimeter results. 2 days in 2012 were unusable due to a large number of obstructions inside the chamber.

**For Proper Equations**

[Click here to download Supplementary material for on-line publication only: GammaPaperDraft3.pdf]

**Figure 1**

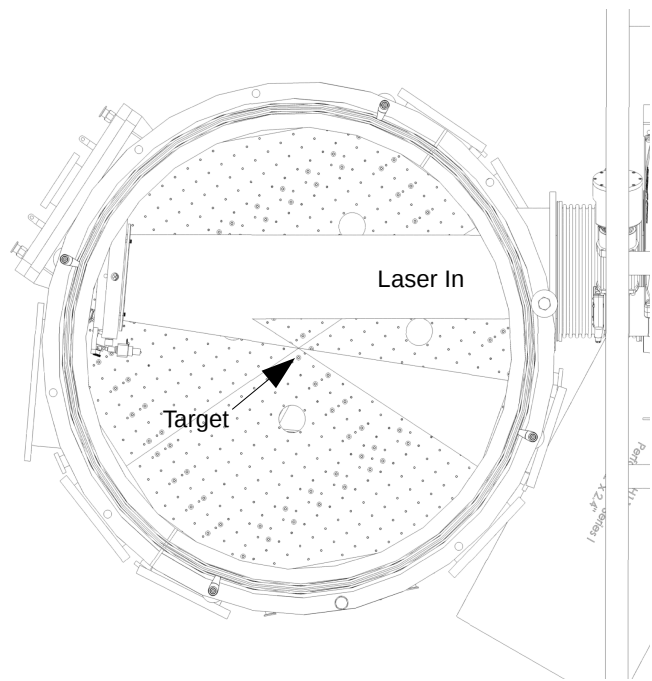 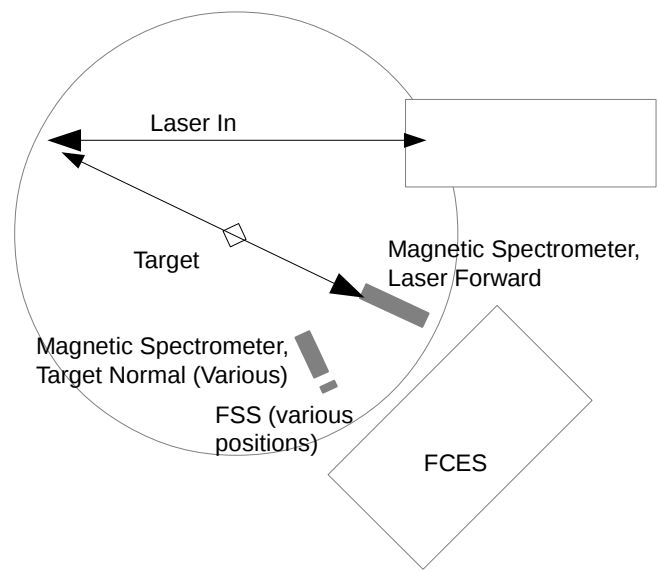

(a) (b)

**Figure 2**

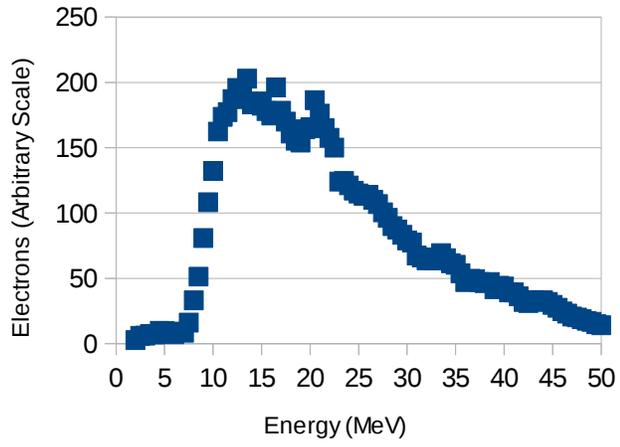 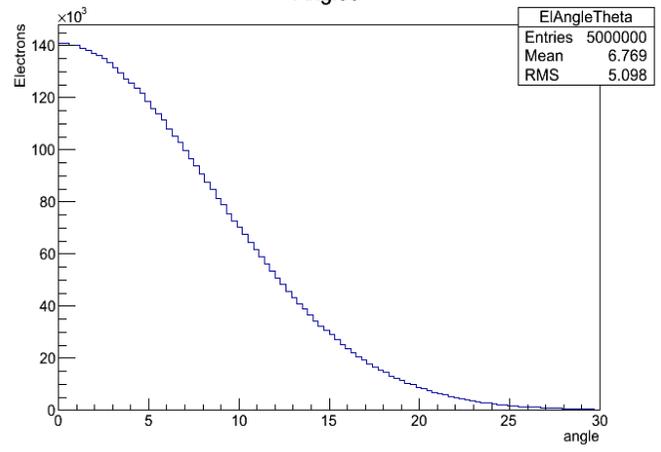

(a)    (b)

**Figure 3**

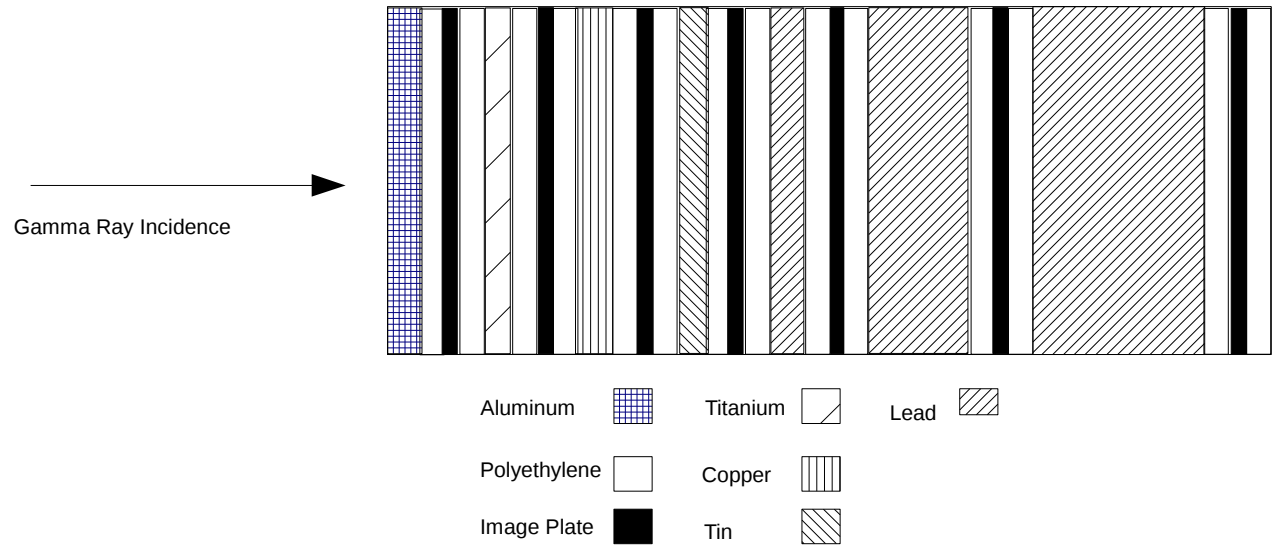

**Figure 4**

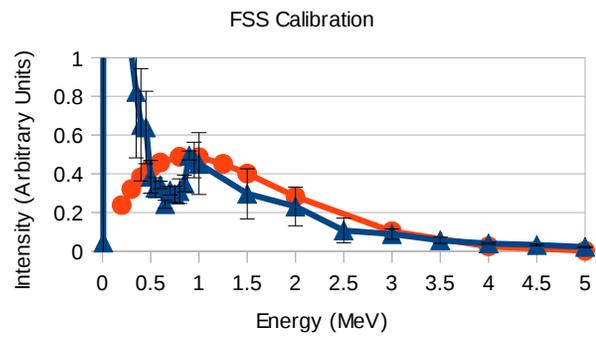

**Figure 5**

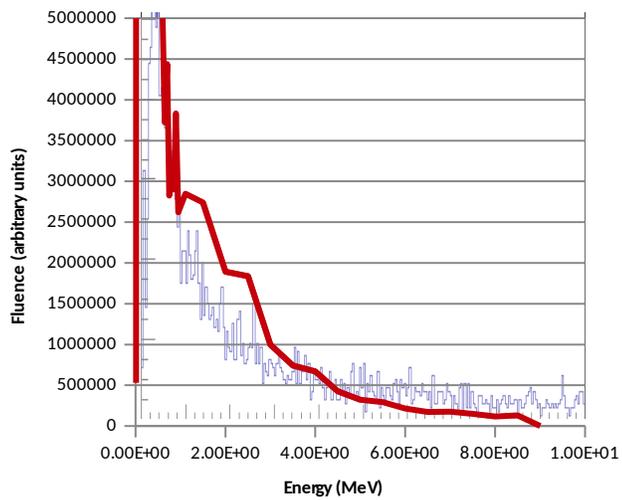

(a)

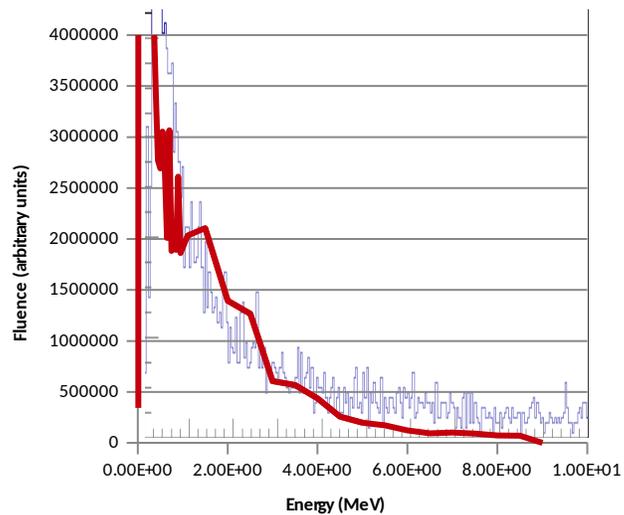

(b)

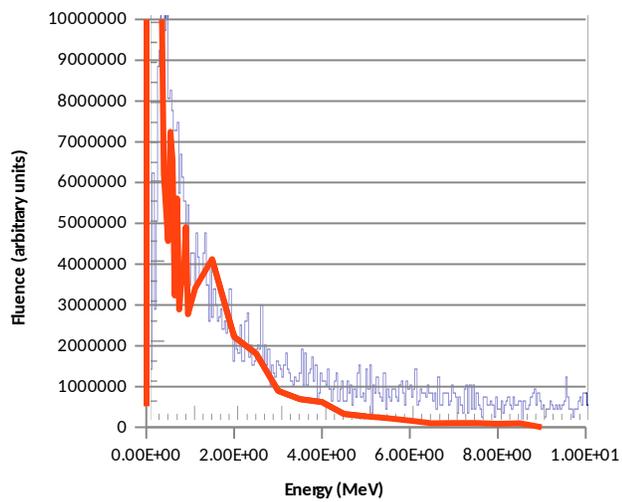

(c)

**Figure 6**

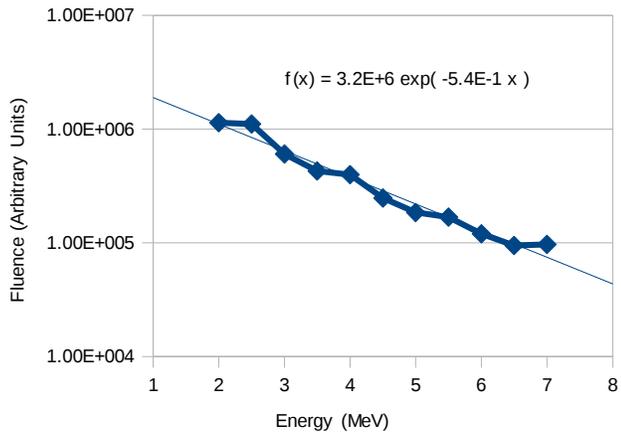

(a)

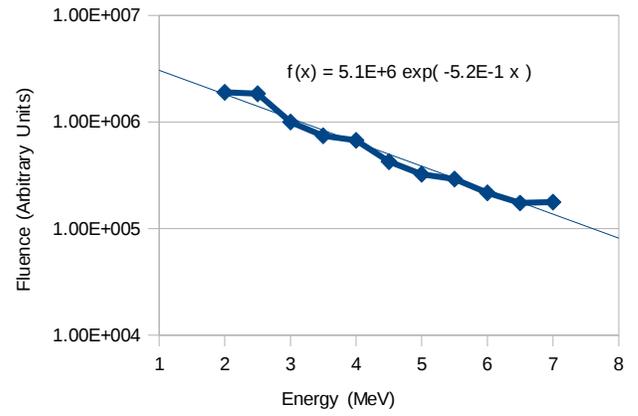

(b)

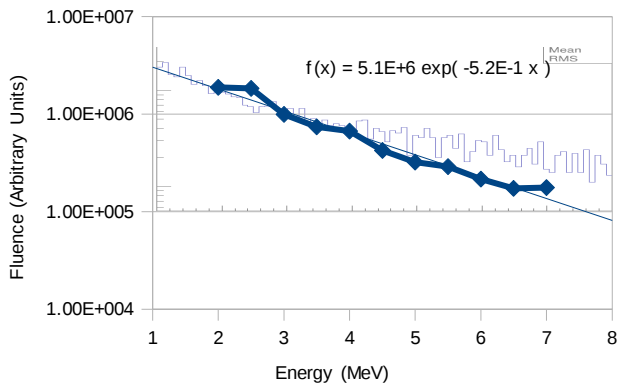

(c)

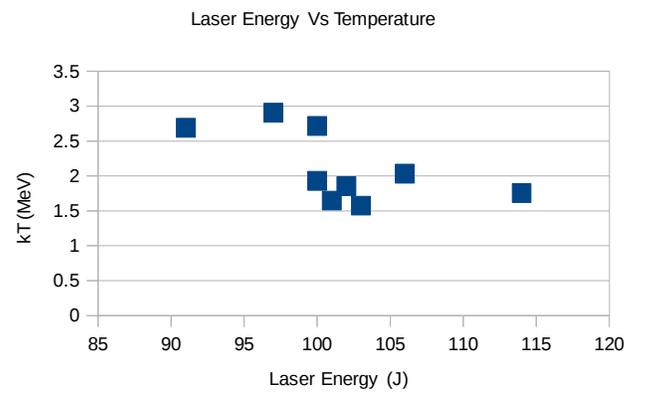

(d)

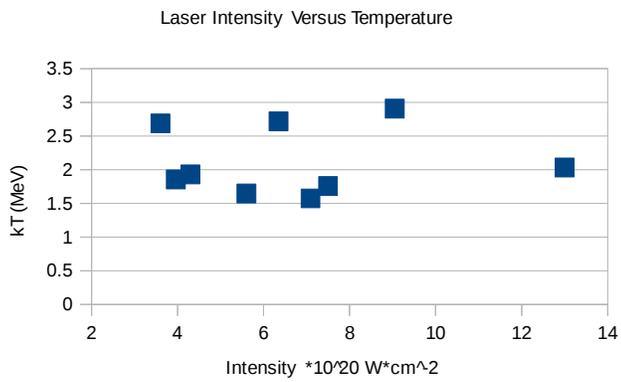

(e)

**Figure 7**

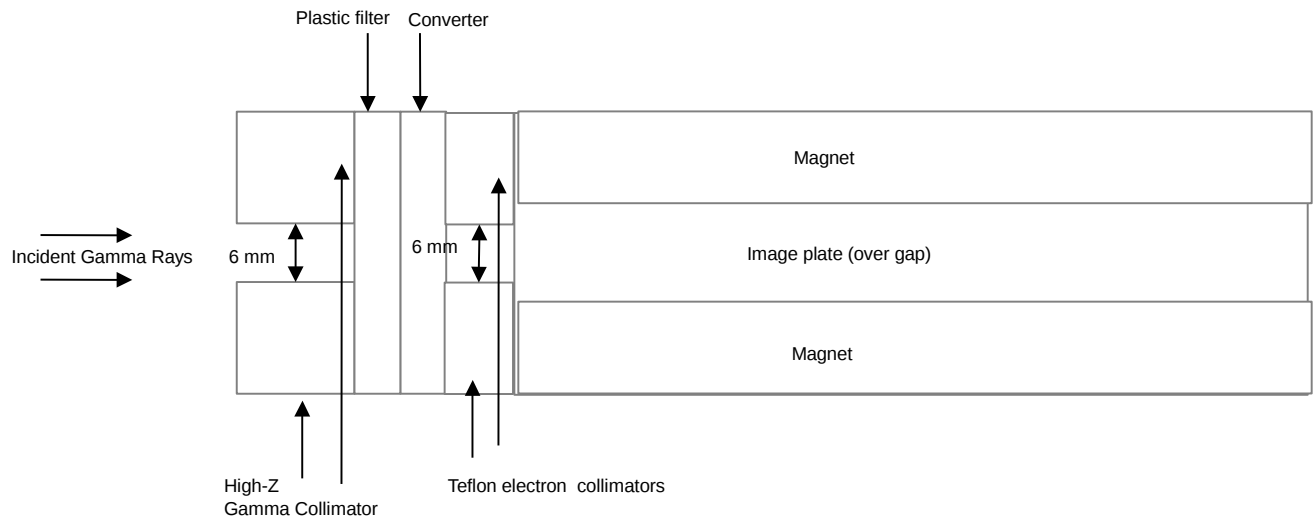

**Figure 8**

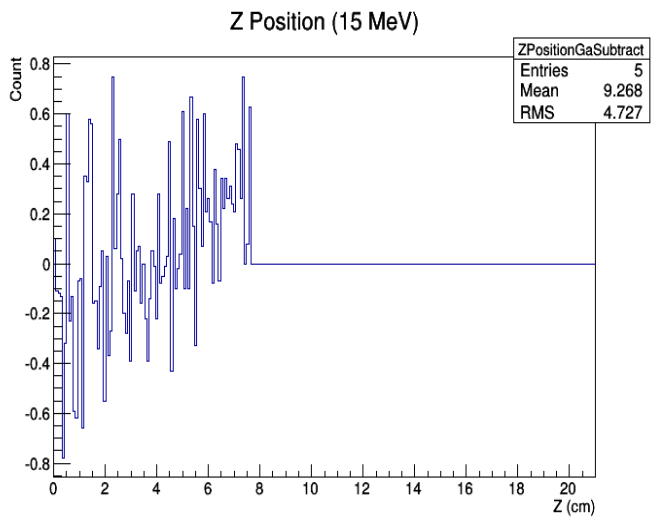

(a)

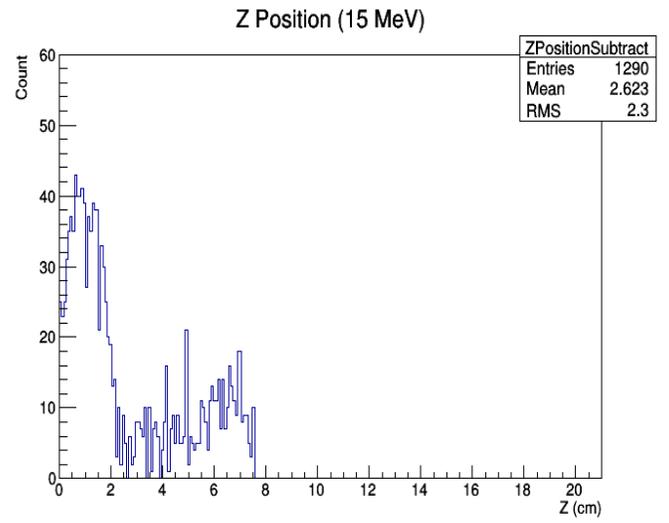

(b)

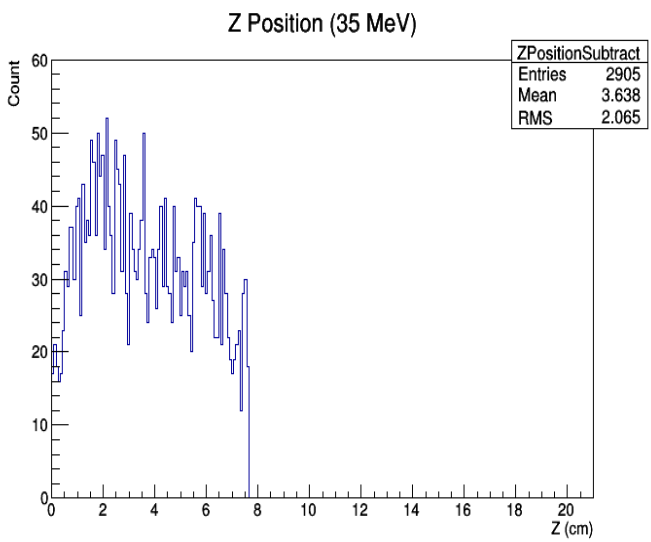

(c)

**Figure 9**

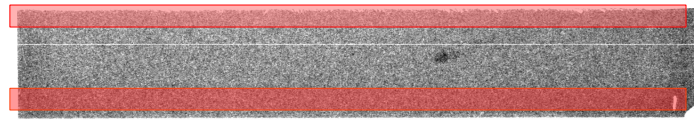

e-

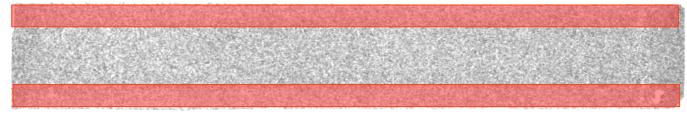

e+

Separate Background Region (shaded regions)

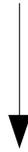

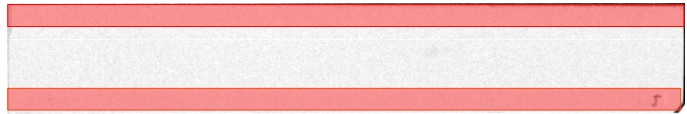

Subtract e+ plate from e- plate, by region

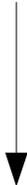

Average background region in pixel columns

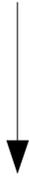

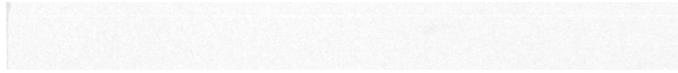

Subtract background from the rest of the image, by pixel column

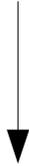

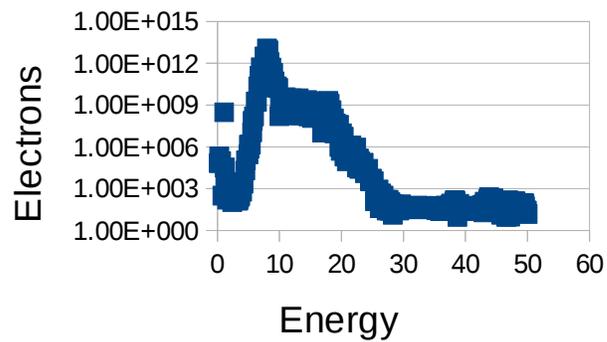

Deconvolve image into Compton electron spectrum

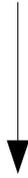
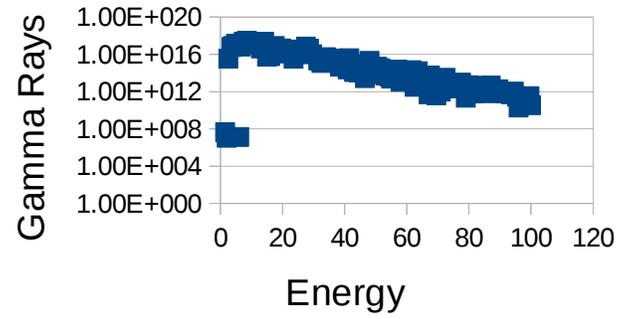

Deconvolve electron spectrum into gamma-ray spectrum

**Figure 10**

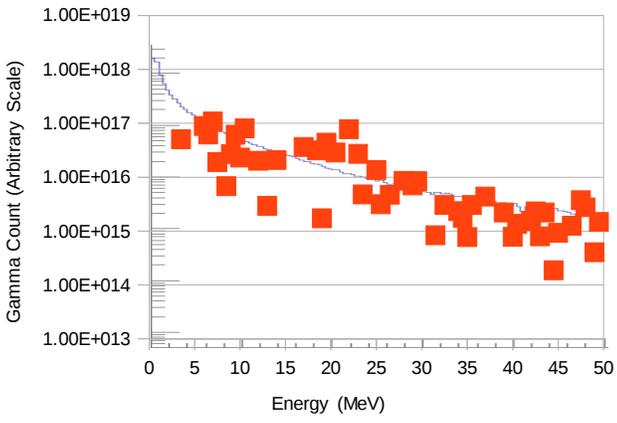

(a)

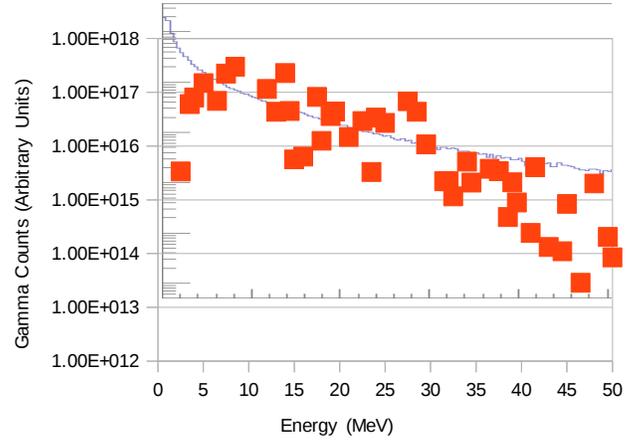

(b)

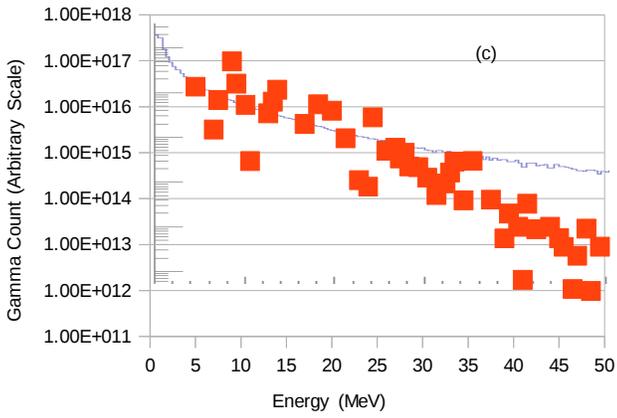

(c)

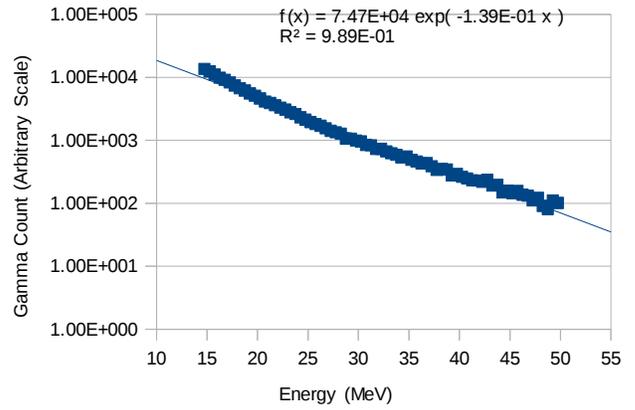

(d)

**Figure 11**

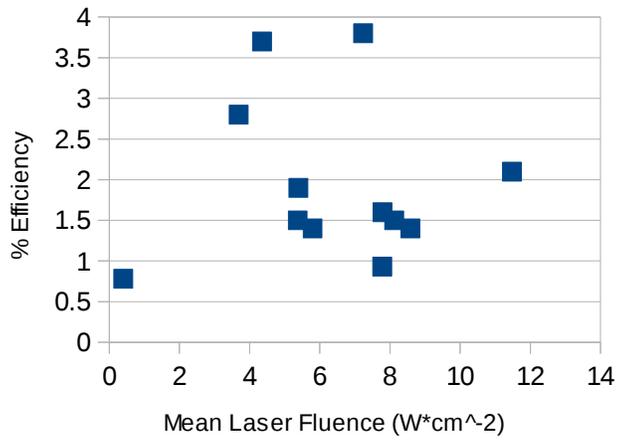
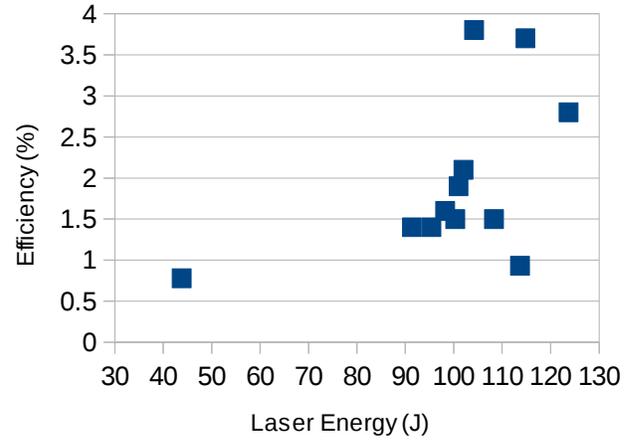

(a)                    (b)

**Figure 12**

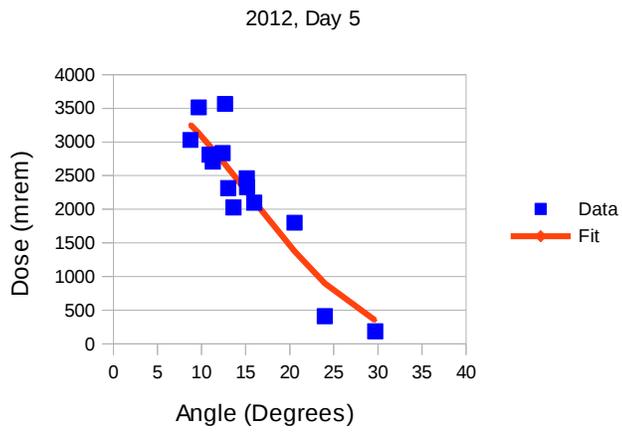

(a)

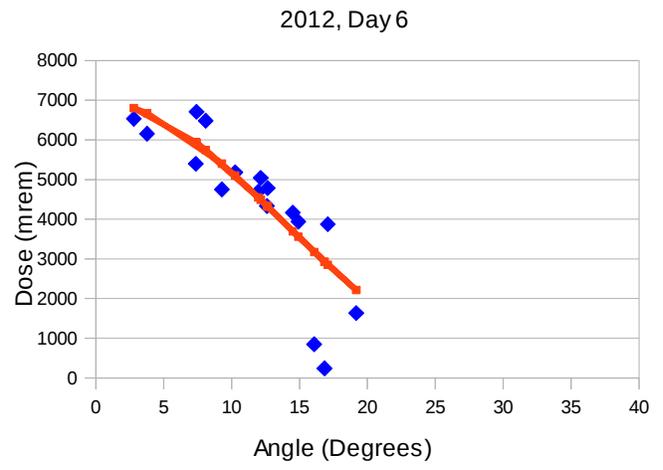

(b)

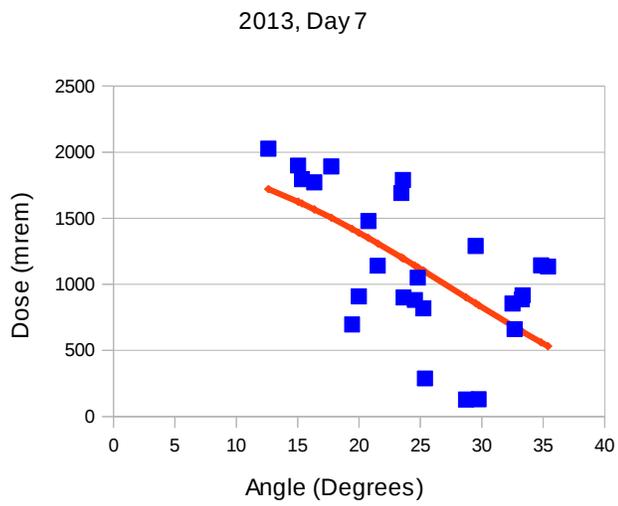

(c)

**Figure 13**

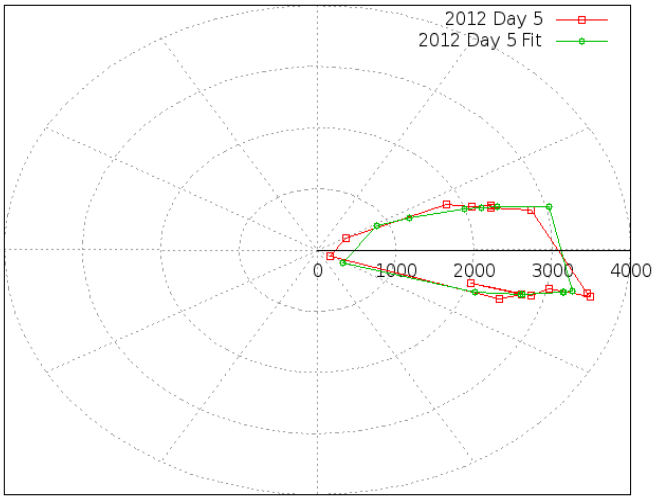

(a)

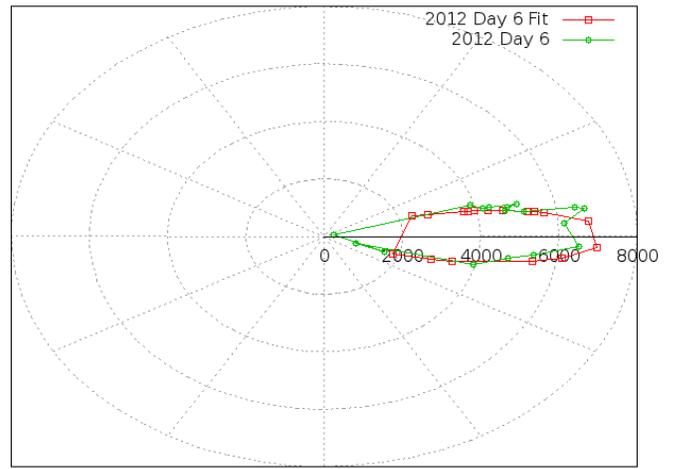

(b)

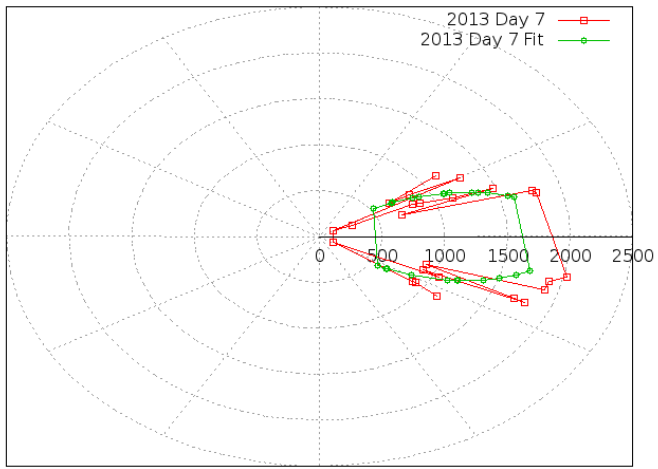

(c)

**Figure 14**

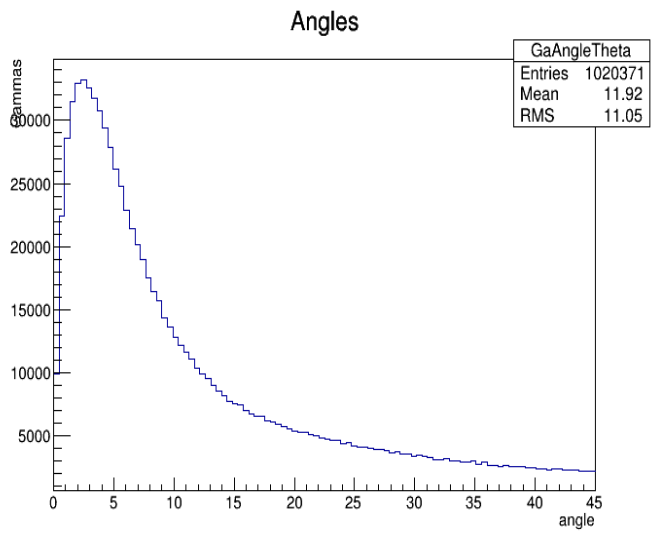